\documentclass[10pt,preprint]{aastex}
\usepackage{epsfig}

\def\p3m{P${}^3$M}
\def\ap3m{AP${}^3$M}
\def\-{{\em{---}}}

\newcommand{\be}{\begin{equation}}
\newcommand{\ba}{\begin{eqnarray}}
\newcommand{\ee}{\end{equation}}
\newcommand{\ea}{\end{eqnarray}}
\newcommand\degrees[1]{\ensuremath{#1^\circ}}

\begin{document}
\title{Two Disk Components from a Gas Rich Disk-Disk Merger }
\author{Chris Brook\altaffilmark{1,2}, 
        Simon Richard\altaffilmark{1,3}, 
        Daisuke Kawata\altaffilmark{4,5}, 
        Hugo Martel\altaffilmark{1}, \& 
        Brad K. Gibson\altaffilmark{6}
\altaffiltext{1}{D\'{e}partement de physique, de g\'enie 
                 physique et d'optique, Universit\'{e} Laval, 
                 Qu\'{e}bec, Qc, G1K 7P4, CANADA}
\altaffiltext{2}{Department of Astronomy, University of Washington, Box 351580,
                 Seattle, WA 98195, USA}
\altaffiltext{3}{Hubert Reeves Fellow}
\altaffiltext{4}{The Observatories of the Carnegie Institution of Washington, 
                 813 Santa Barbara St., Pasadena, CA 91101}
\altaffiltext{5}{Swinburne University, Hawthorn VIC 3122, Australia}
\altaffiltext{6}{Centre for Astrophysics, University of Central 
                 Lancashire, Preston, PR1, 2HE, UK}}

\begin{abstract}
We employ N-body, smoothed particle hydrodynamical simulations, including 
detailed treatment of chemical enrichment, to follow a gas-rich merger which 
results in a galaxy with disk morphology. We trace the kinematic, structural, 
and chemical properties of stars formed before, during, and after the merger. 
We show that such a merger produces two exponential disk components, with the 
older, hotter component having a scale-length 20\% larger than the 
later-forming, 
cold disk. Rapid star formation  during the merger quickly enriches 
the protogalactic gas reservoir, resulting in high metallicities of the 
forming stars. These stars form from gas largely polluted by Type II 
supernovae, which form rapidly in the merger-induced starburst. After the 
merger, a thin disk forms from gas which has had time to be polluted by 
Type Ia supernovae. Abundance trends are plotted, and we examine the proposal 
that increased star formation during gas-rich mergers may explain the high 
$\alpha$-to-iron abundance ratios which exist in the relatively 
high-metallicity thick disk component of the Milky Way.  
\end{abstract}


\section{Introduction}

That major mergers of two gas-rich disk galaxies can sporn a disk galaxy 
was shown in \cite{springel}.  Disk galaxies can result from such events 
for a  large range of mass ratios, orbits, and rotational velocities of 
progenitor galaxies (\citealt{robertson06}).  The discovery that thick 
disks are common, and perhaps even ubiquitous in disk galaxies 
\citep{dalcanton,mould}, and their old age  means that the thick disk holds 
important clues to the formation of such galaxies. It also implies that 
disk galaxies forming from a major merger  should have a thick disk 
component.  Examining Figure~2 of \cite{robertson06}, it is clear that 
a relatively hot (thick) disk component  will naturally form in such 
mergers, motivating our desire to further investigate such events. 
In general terms, mergers were presumably increasingly gas-rich at 
increasingly early epochs. The possibility that gas-rich mergers play 
an essential role in disk galaxy formation has gained momentum in recent 
years. \cite{brook04b,brook05} suggest that the thick disk is formed in a 
high-redshift, gas-rich merger epoch, characteristic of a $\Lambda$CDM 
Universe, during which the galaxy is born. High-redshift mergers being 
gas-rich is also consistent with other constraints including a low-mass, 
metal-poor stellar halo (\citealt{brook04a}), the chemical abundance 
patterns of the stellar halo (\citealt{renda,font}), and the angular 
momentum of disks (e.g. \citealt{governato}). For a review of these 
issues, see \cite{brook06}.  

The ratio of $\alpha$ element to iron abundances has been used as a ``clock''  
to infer different formation time-scales for different galaxy types, as well 
as for components of the Milky Way. Type II supernovae (SNe II), which trace 
star formation closely due the short life-spans of their high-mass progenitor 
stars, produce large quantities of oxygen, magnesium, silicon --- 
the so-called 
$\alpha$ elements. By contrast, iron is produced predominately in  Type Ia 
supernovae (SNe Ia), whose explosions are delayed. Hence, the high 
$\alpha$-to-iron ratio characteristic of elliptical galaxies is interpreted
as implying a rapid formation, while the low $\alpha$-to-iron ratio of
disks and dwarf galaxies implies a protracted formation. In our Galaxy,
the low metallicity stellar halo, with peak iron abundance
$\rm[Fe/H]\sim-1.5$ has a ratio $\rm[\alpha/Fe]\sim0.4$, while the
relatively metal rich thick disk (peak $\rm[Fe/H]\sim-0.6$) also has
enhanced $\alpha$ elements compared with solar abundances.
Recently, several studies (e.g. \citealt{bensby,reddy}) have highlighted
the difference in the $\alpha$-to-iron element abundance ratio between the
hot and cold disk components of the Milky Way, providing  a clue to the
processes involved in the formation of the components. In \cite{brook05},
we suggested that  high star formation rates in gas-rich mergers result 
in increased metallicities with high $\alpha$ element abundances. Such 
events may be central in forming the thick disk component of the Milky 
Way.  Here, we  examine in detail a single  isolated example of such a
merger, and follow the [$\alpha$/Fe] ratio, and how it evolves as the 
metallicity of the final galaxy increases during the merger event. 

 Our study shows that two disk components naturally form in a gas-rich
merger, and we examine structural and  kinematic features of the resulting
galaxy. We show that the starburst which accompanies this merger has clear
chemical signatures. We analyze and compare abundances in 
two specific groups of stars: {\it merger stars}, which
are formed before and during the merger, and
{\it disk stars}, those forming in the relatively quiescent
period after the merger. The {\it starburst} which accompanies the merger
results in rapid chemical enrichment with the large number of SNe~II forming,
ensuring high $\alpha$ element abundances. We discuss implications for the
end products of gas-rich merger events, and the importance of such events
in the formation of disk galaxies. 
This merger simulation will form part of a larger suite of simulations which
explore the kinematic, structural, and chemical properties of gas-rich
mergers remnants with various values of progenitor mass ratio, rotational
velocities, and impact angles.

\section{Simulation Details and Results}

We simulate the merger using GCD+, which self-consistently models
the effects of gravity, gas dynamics, radiative cooling, and star
formation.  We give an outline of the code here, whilst full details
are found in \cite{kawata}. 
GCD+ is a Tree/SPH algorithm that
includes SNe~Ia and SNe~II feedback, and traces the lifetimes of individual
stars, enabling us to  monitor the
chemical enrichment history of our simulated galaxies. Star formation occurs
in a convergent gas velocity field where gas density is greater than a
critical density, $\rho_{\rm crit}= 2 \times 10^{-25}{\rm g/cm}^3$. The star
formation rate (SFR) of eligible gas particles is then
$d\rho_*/dt=d\rho_g/dt=c_*\rho_g/t_g$
where $c_*=0.5$ is a dimensionless star formation efficiency,
and $t_g$ is the dynamical time. This formula corresponds to the
Schmidt law: SFR $\propto\rho^{1.5}$. The mass, energy, and heavy
elements are smoothed over the neighboring gas particles using the SPH
smoothing kernel.
We assume that $10^{51}{\rm ergs}$ is fed back as thermal energy from each
SNe. Gas within the SPH smoothing kernel of SNe~II explosions is prevented
from cooling, creating an adiabatic phase for gas heated by such SNe. This
adiabatic phase is assumed to last for the lifetime of the lowest mass star
that ends as a SN II, i.e., the lifetime of an $8\,{\rm M}_\odot$ 
star (100 Myr). This is
similar to a model presented in \cite{thacker}. For SNe~II, the
metallicity-dependent stellar yields of \cite{ww} are adopted. For
low- and intermediate-mass stars, we use the stellar yields of
\cite{vanden}. We adopt SNe Ia model of \cite{kobayashi}, and the
yields of \cite{iwamoto}.

The initial conditions are two galaxies with exponential gas disks,
embedded in  dark matter halos. These are created using GalactICS
(\citealt{dubinski}), and are essentially stable in the that their
density profiles, potential, and velocity ellipsoids will not change
significantly when individual galaxies are evolved. GalactICS uses the
lowered Evans model for the dark matter halo, which leads to a
constant-density core. Numerical simulations suggest that halos have
a cuspy central density profile
\citep{nfw96,nfw97,mooreetal99,ghignaetal00,js00,klypinetal01}. However,
these results are in conflict with many observations, so we prefer to use
a constant-density core. In any case, our results should not be
sensitive to the details of the central density profile, since
the core contains only a small fraction of the total mass.

The larger galaxy has total mass of
$5\times10^{11}{\rm M}_\odot$. The mass ratio is 2:1, disk scale lengths
are 4.5 and 3.1 kpc, and each has baryon fraction of 17$\%$. Each galaxy
consists of 40,000 baryonic and  100,000 dark matter particles. The small
galaxy approaches with its orbital angular momentum vector inclined
\degrees{17} to that of the large galaxy. Both galaxies have prograde
rotation with respect to the angular momentum of the system, the orbital
energy of the system is 1.7$\times 10^{44}$ ergs. The system is bound,
and the spin parameter (ratio of orbital energy to binding energy) is
$\lambda=0.04$.
The gas disks evolve and form stars prior to the merger, and have gas
fractions of $f_g\equiv M_{\rm gas}/(M_{\rm gas}+M_{\rm stars})=0.91$ at
the time of the merger. Gas is given an initial metallicity of
$\log(Z/Z_{\odot})=-4$, and $\rm[\alpha/Fe]=0.35$. We follow the
simulation for 1.5 Gyrs.

\begin{figure}[t]
\epsfig{file=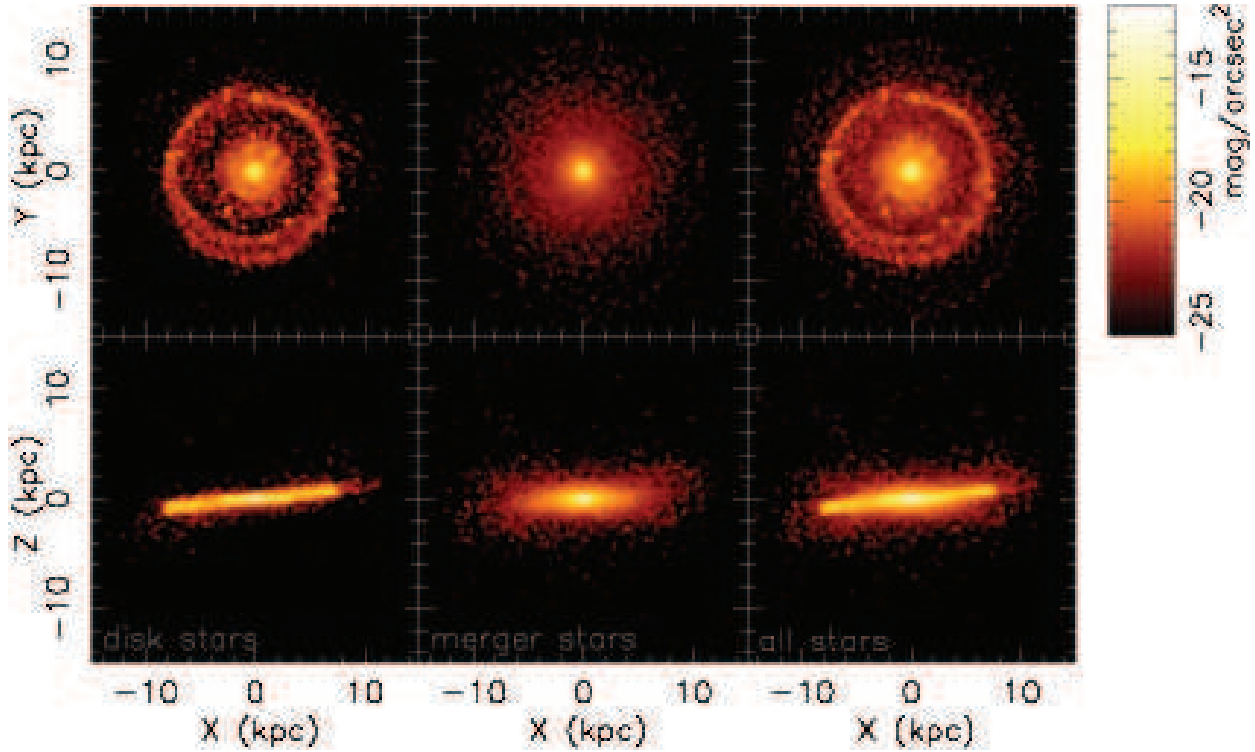,width=6.5in}
\caption{B-band Luminosity for final distribution, shown face on 
(upper panels) and edge-on (lower panels) of all stars ({\it right}), 
{\it merger stars}, (stars formed before and during the merger, 
{\it middle}), and {\it{disk stars}}, (formed after the merger, 
{\it left}). $X$, $Y$, $Z$ are cartesian coordinates, with $Z$ being along 
the axis of rotation.}
\label{final}
\end{figure}

\begin{figure}
\epsfig{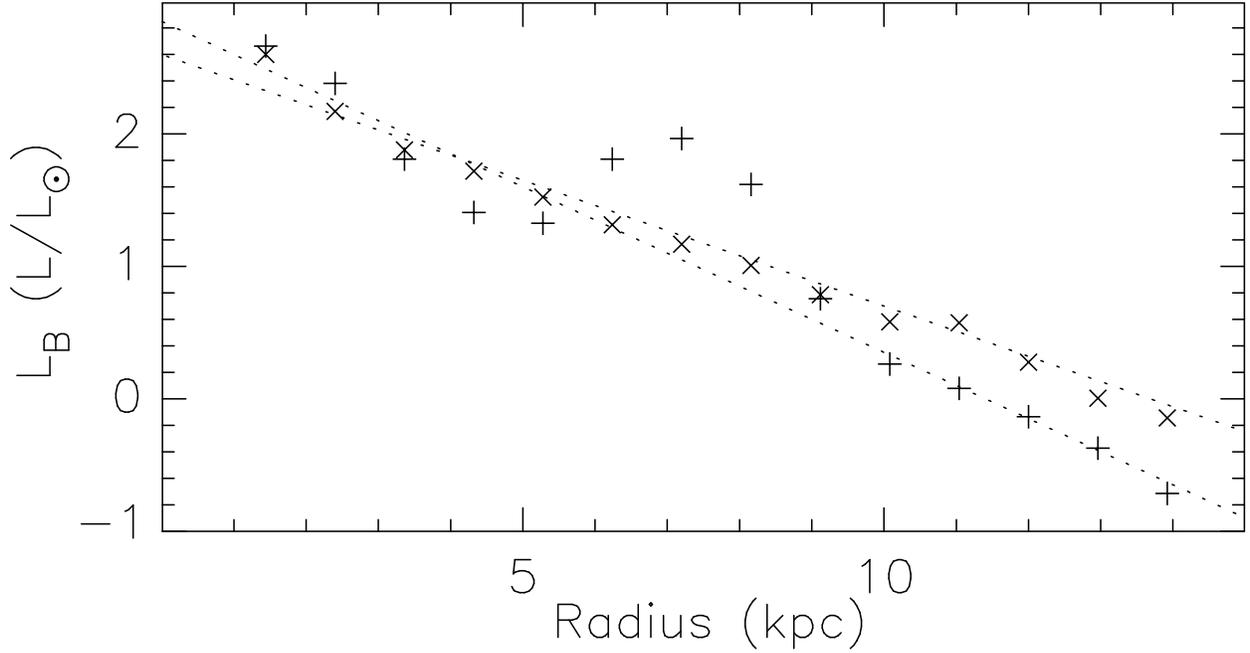}
\caption{B-band Luminosity profiles of merger stars ($\times$ symbols) 
and disk stars ($+$ symbols). Dotted lines are fits between 2.5 and
15 kpc, and indicate scale-lengths of 5.1 and 4.1 kpc respectively
(ignoring a dense region associated with a ring in the disk stars).}
\label{struct}
\end{figure}

The gas-rich merger results in a final galaxy with disk morphology. 
Figure~\ref{final} shows the B-band luminosity map of the resultant galaxy 
after 1.5 Gyrs of the simulation,  where
we employ the simple stellar population of \cite{kodama}.
Shown, both face on (upper panels) and edge on (lower panels) are all 
stars (right panel), what we will call {\it{merger stars}}, which are those 
formed before and during the merger (middle panel), and {\it{disk stars}}, 
which are those that form after the merger (left panel). From the face-on 
view, it is evident that the  remnant of this particular merger simulation 
is a ring galaxy, which indicates that prograde gas-rich disk-disk mergers 
can produce ring galaxies. We find this especially interesting in light of 
\cite{Lavery}, who found that the incidence of rings increases rapidly with 
redshift. 
We will study this further in our future papers.

The B-band luminosity profiles are shown in Figure~\ref{struct}, and used 
to determine scale-lengths of 5.1 and 4.1 kpc respectively for the 
merger and disk stars. 
A bulge component, within the inner $\sim3{\rm kpc}$, 
is apparent from the surface 
density profile, which is best fit with a bulge, disk, and thick disk.
This is consistent with the merger remnant depicted in Figure~2 of
\citet{robertson06}. We note that our idealized initial disks do not have
spheroid stellar components. Any stars in such component would end up in a
spheroidal component in the final galaxy, and thus the idealized initial
conditions are partially responsible for the lack of a significant
spheroidal component in the final galaxy. Our results remain valid so
long as any initial spheroid component is of low enough mass not to effect
the dynamics of the merger. 

\begin{figure}
\epsfig{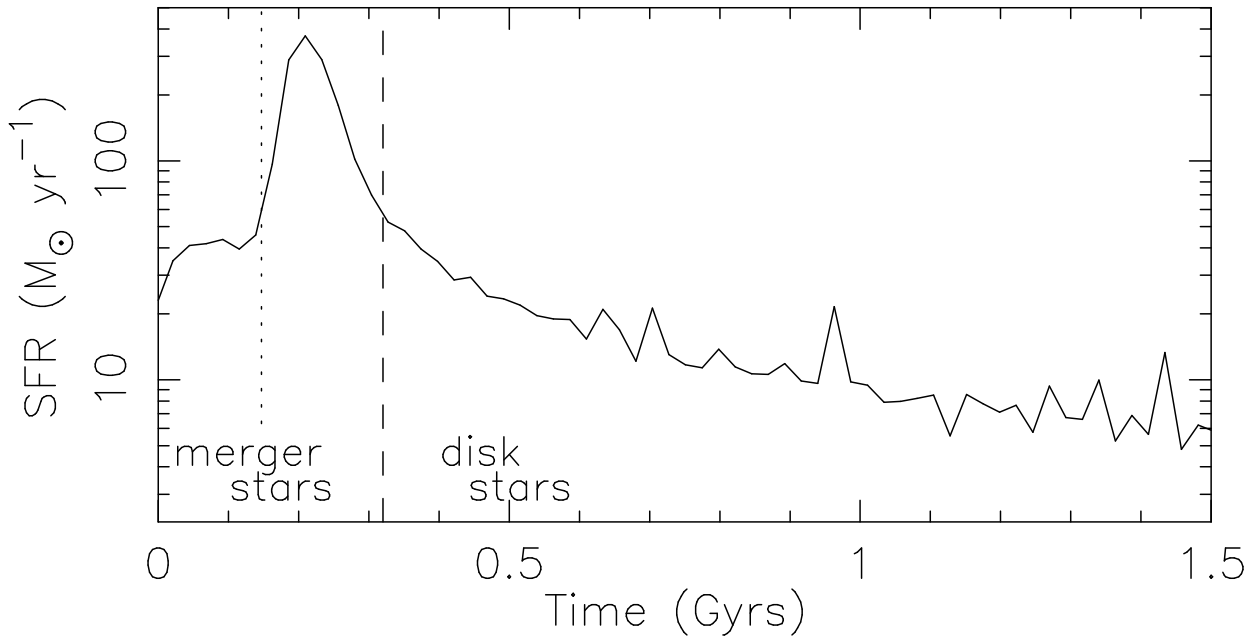}
\caption{Star formation rate (M$_\odot$/yr) against time in Gyrs. The 
starburst during the merger has a peak SFR of 380 M$_\odot$/yr. The 
dashed line shows the division between what we have termed
{\it merger stars} and {\it disk stars}. The dotted line indicates where
the merger begins.} 
\label{sfr}
\end{figure}

\begin{figure}[t]
\hskip0.5in
\epsfig{file=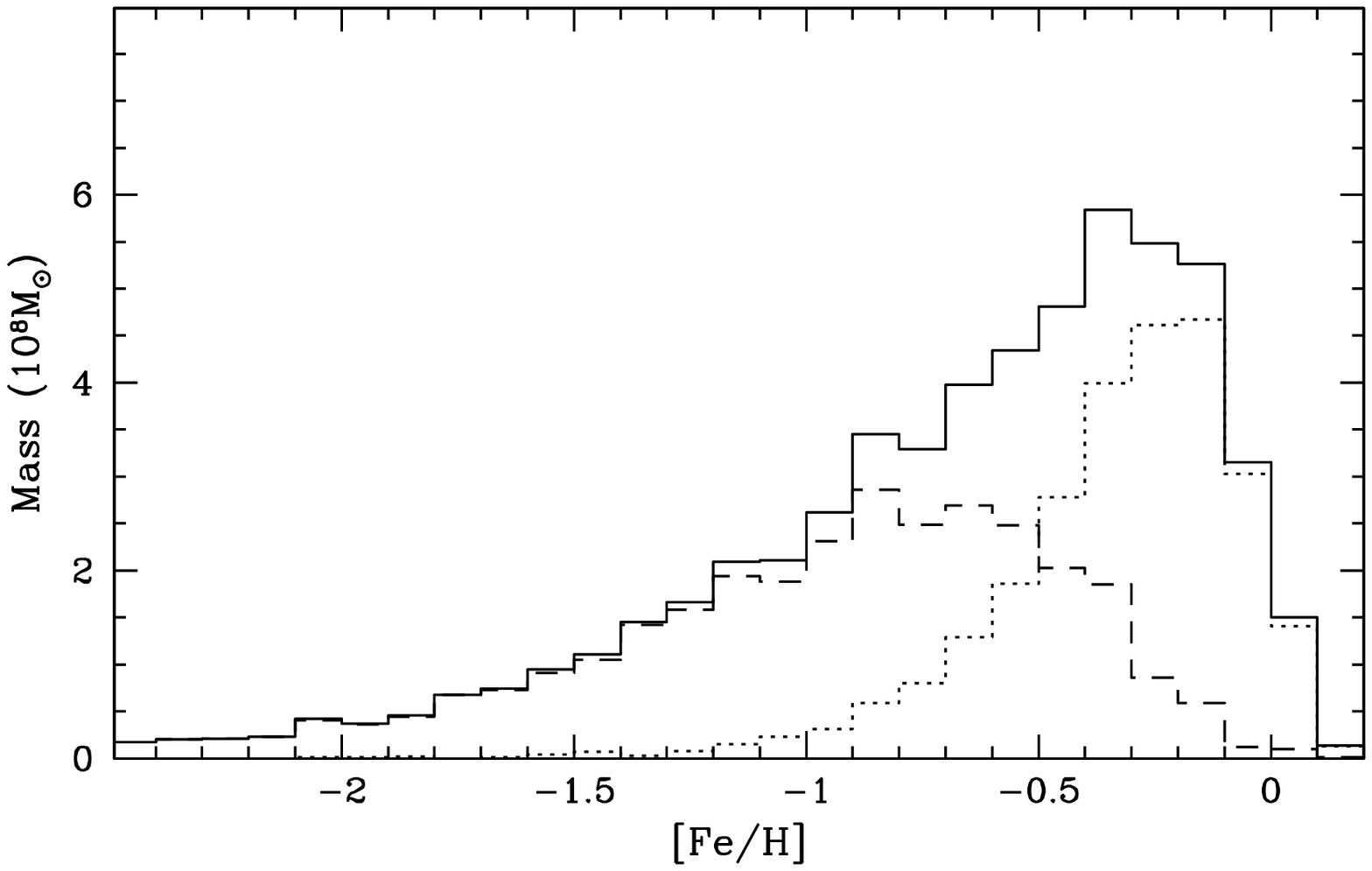,width=5.5in}
\caption{The metallicity distribution function for stars in the region
$4<R_{XY}<10$ and $|Z|<1{\rm kpc}$ 
in the final galaxy. All stars are plotted as
a solid line, and broken up into disk stars (dotted line), 
and merger stars (dashed line).} 
\label{MDF}
\end{figure}

The star formation rate, Figure~\ref{sfr}, shows a starburst which peaks
at around $380\,{\rm M}_\odot{\rm yr}^{-1}$, during the merger. 
The end of this
starburst is used to divide stars  into merger and disk stars.
Prior to the merger, star formation is around 30 M$_\odot$yr$^{-1}$. After
the merger event, the star formation rate drops below
$10{\rm M}_\odot{\rm yr}^{-1}$
after around a Gyr, although this is affected by our idealized initial
conditions that assumes a dense gas disk. The mass of stars born before,
during, and after the merger, are $6.3\times10^9$, $33\times 10^9$, and
$19\times10^9{\rm M}_\odot$ respectively. The merger stars have a
thicker morphology than disk stars. In the disk region,
$4<R_{XY}<10\,{\rm kpc}$ and $|Z|<1\,{\rm kpc}$, the merger and
disk stars have total stellar mass of 3.5$\times 10^9{\rm M}_\odot$,
and $4.4\times 10^9{\rm M}_\odot$ respectively. These two components also
separate when we consider their  metallicity distribution function.
Figure~\ref{MDF} shows the metallicity distribution function for merger
and disk stars. The merger stars have
a peak metallicity of around $\rm[Fe/H]\sim-0.8$, while the disk stars
peak at $\rm[Fe/H]\sim-0.2$. The long low-metallicity tail of the merger
component is due to a difference in metallicity between merger stars
born before and during the merger-induced starburst.   

\begin{figure}
\hskip0.8in
\epsfig{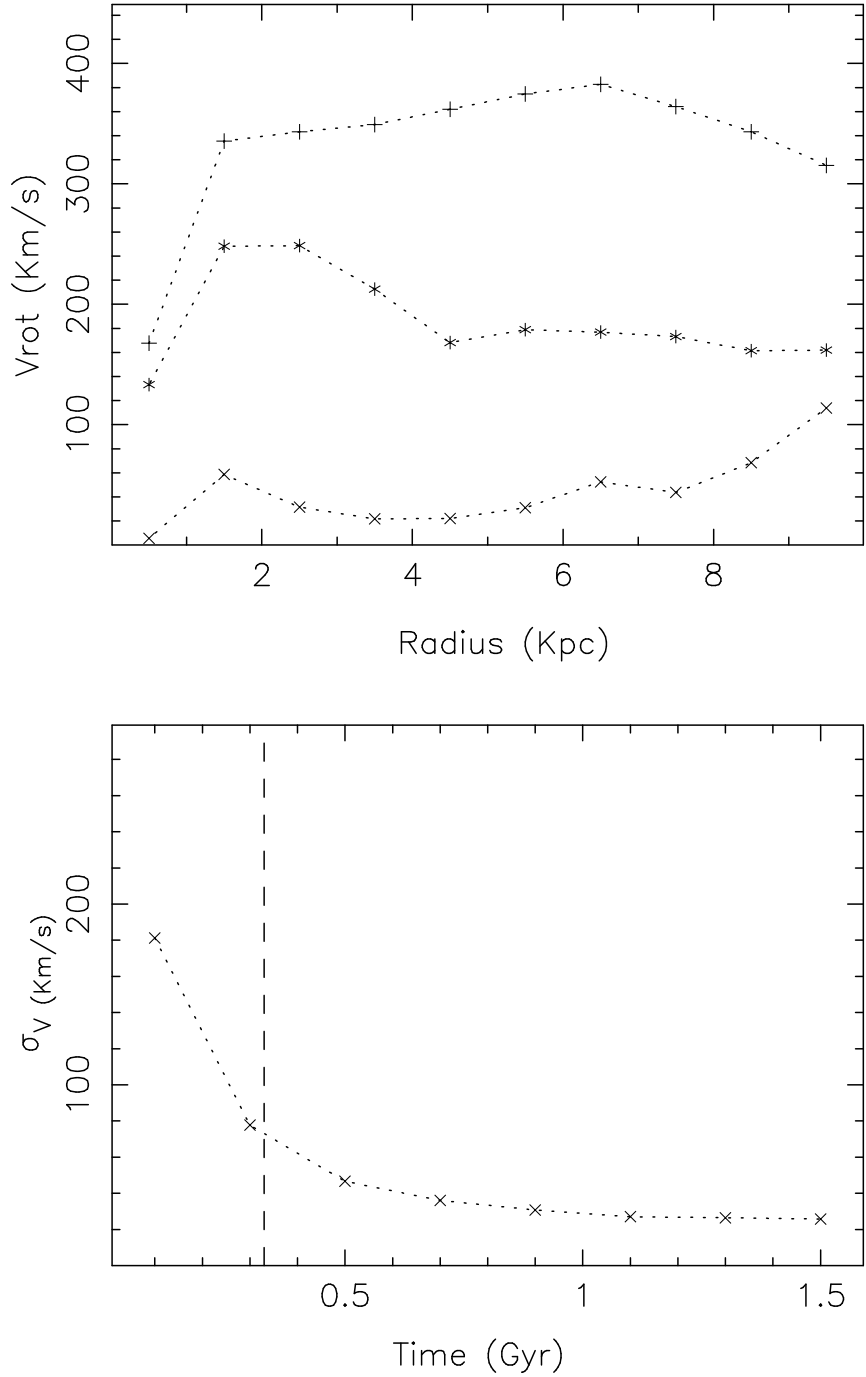}
\caption{{\it Top Panel:} Rotational velocity plotted against radius for 
merger stars ($\times$ symbols) and disk stars ($+$ symbols). 
When a counter-rotating component is ignored, the remaining merger stars
are shown as $*$ symbols. {\it Bottom Panel:} Velocity dispersion in the 
direction of rotation versus formation time of final stars. The dashed line
corresponds to the end of the merger, from Figure~\ref{sfr}.}
\label{vel}
\end{figure}

The top panel of Figure~\ref{vel} shows the rotation velocity of
merger and disk stars. Not surprisingly, the merger stars
have significantly slower rotation. It is worth mentioning that 
a part of merger stars are counter-rotating. We also measure the 
rotation excluding the counter-rotating stars, and found that their 
rotation is still much lower than the disk stars.
The timing of our division between merger and disk stars is 
consistent with the change in velocity dispersion, as seen
in the bottom panel of Figure~\ref{vel} which plots dispersion of
rotational velocity versus formation time of the stars. 

The chemical abundances of the merger simulation are explored in
Figure~\ref{alpha}. The average abundances of O, Mg, and Si are used to 
calculate [$\alpha$/Fe], and we take a slice through the galaxy at 
$|z|<1{\rm kpc}$. The [$\alpha$/Fe] versus radius plot (top left panel) shows 
the merger stars ($\times$ symbols) have $\rm[\alpha/Fe]\sim0.35$ 
independent of radius (although the inner regions, associated with a bulge, 
is a little lower), whilst the later-forming disk stars ($+$ symbols) have 
a value $\sim0.15\,{\rm dex}$ lower. The plot of [$\alpha$/Fe] versus 
[Fe/H] in 
the top middle panel is interesting, and motivated this letter.  
The merger stars maintain a higher ratio, 
[$\alpha$/Fe], even when their [Fe/H] has increased to solar values, 
whilst the disk stars which are formed with values of 
$\rm[Fe/H]\sim-0.5$ do so with relatively low values of [$\alpha$/Fe]. 
Examination of the bottom left panel, where we plot the evolution of 
[$\alpha$/Fe] with time,  helps to explain this result. The first 
stars have $\rm[\alpha/Fe]\sim0.35$, close to the initial condition, and 
their ratio drops prior to the merger. The value of [$\alpha$/Fe] 
{\it increases} during the time of the starburst, as indicated by the jump 
from the second to the third points, before decreasing to the value found 
in the disk. During the starburst, associated large numbers of SNe~II
 provide a supply of $\alpha$ elements which allows the maintenance of 
a high ratio of [$\alpha$/Fe], even as the iron content of the gas reservoir
is increased (bottom middle panel). After the starburst, the pollution of
SN~Ia becomes effective, and along with the contribution  of abundances of
low- and intermediate- mass stars, the value of [$\alpha$/Fe] is maintained
at a constant value.
Further information on  disk galaxy formation is found in vertical
abundance gradients.
The upper right panel of Figure~\ref{alpha} shows that neither merger
nor disk stars have vertical  gradient of [$\alpha$/Fe], while
disk stars show a slight decrease in [Fe/H] at larger vertical  height.

\section{Discussion}

\begin{figure*}[t]
\epsfig{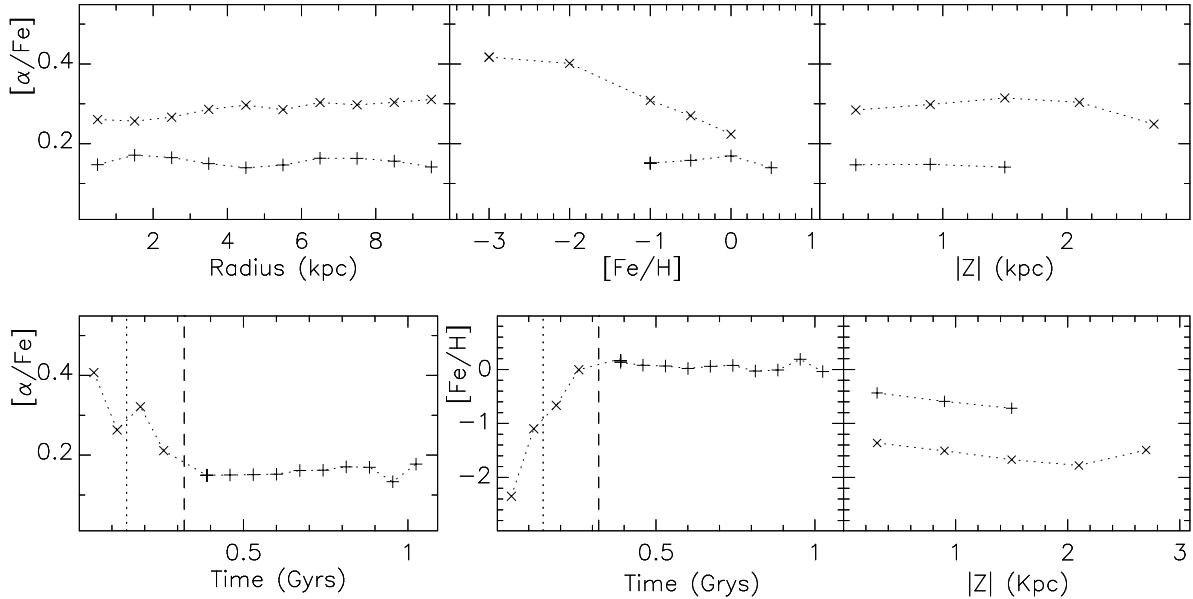}
\caption{In all plots, merger and disk stars use $\times$
and $+$ symbols respectively. {\it Upper panels}, [$\alpha$/Fe]
versus radius ({\it left}), [Fe/H] ({\it middle}), and
vertical height (for stars with $4<R_{XY}<10$, {\it right}). 
{\it Lower left panel}, [$\alpha$/Fe] versus time.
{\it Lower middle panel},
[Fe/H] versus time. 
{\it Lower right panel}, [Fe/H] versus
vertical height (for stars with $4<R_{XY}<10$).}
\label{alpha}
\end{figure*}

In  gas-rich mergers, two disk components naturally emerge. A thick disk
consists of stars born before and during the merger, although we note that
stars born before the merger can also end up in the stellar halo, which they
did in the simulation of \cite{springel}. A thin disk forms rapidly at the
end of the merger. The high star formation triggered in such events can also
help explain  the chemical abundances of the stars. The starburst results
in rapid increase in metallicity (Fig.~\ref{alpha}, bottom middle panel),
with the short star formation timescale ensuring that stars formed are
enriched largely by SNe~II, and hence merger stars maintain a higher
[$\alpha$/Fe], even when their [Fe/H] has increased to solar values. The
stars which form after this starburst epoch are formed with low  values of
[$\alpha$/Fe], even those formed with low $\rm[Fe/H]\sim-0.5$, and this
quiescent epoch means that the stars form in a thin disk, with low velocity
dispersion (Fig.~\ref{vel}, bottom panel).  
 
In \citet{brook04b}, we found that the thick disk scale-length of our 
simulated galaxy was shorter than that of the thin disk. The observations 
of \cite{yoach2} find that thick disk scale-lengths are systematically 
larger than those of thin disks. In this merger simulation, the hot merger 
star population has an exponential profile with scale-length larger than 
that of the later-forming disk star population.  This may favor a 
significant gas-rich merger being a feature of disk galaxy formation. 
Further simulations will determine whether  old, hot disks  with scale-length 
larger than those of young cold disks  result for a wide range of merger 
parameters. Caution is required in interpreting the result, as subsequent 
``inside-out'' thin disk growth through  gas infall (not present in the
current simulation) may increase the thin disk scale-length \citep{brook06b}.
\cite{yoach2} further find that   low-mass disk galaxies have larger
thick:thin disk mass ratios. They interpret this as evidence for the
formation of the thick disk by direct accretion of stars, as progenitors of
lower mass galaxies will more easily expel their gas  from their small
potential well prior to the merger. Yet observations suggest that low mass
galaxies are in fact more gas-rich, both at low redshift
(e.g. \citealt{schombert}) and high redshift (\citealt{erb}), favoring
gas-rich mergers as an interpretation of the higher thick:thin disk mass ratio
in low mass galaxies. Also, the  growth of the thin disk in small galaxies
is perhaps regulated by their low densities (\citealt{dalcanton06}) and
supernovae feedback, i.e. the high thick to thin disk ratio of low-mass
galaxies may be the result of less  growth of a thin disk.

Our idealized study does not include cold accretion from the intergalactic 
medium and infall processes, which are important in the birth of disk 
galaxies. Further, in the hierarchical structure formation scenario, a range 
of merger histories exists, and to explain the predominance of thick disk 
components in disk galaxies, one cannot rely on a single merger event. Rather, 
a few or several significant gas-rich merger events are likely to occur at an 
early epoch  in the formation of a disk galaxy (\citealt{brook05}). A cold 
mode of accretion from filamentary structures also occurs in this scenario. 
But our study supports the view that the violent accretion of gas-rich 
galaxies is central in  producing thick disk properties. By ignoring cold 
accretion, our simplified  study highlights the effect of the starburst 
associated with such mergers on the chemical abundances of the forming stars, 
in particular the high $\alpha$ element abundances at high metallicities, 
and vertical abundance gradients. It is yet to be shown that cold accretion 
alone, perhaps resulting in a thin disk which is heated by an infalling 
satellite, or the dispersion of large star clusters 
(\citealt{kroupa,elmegreen}), can reproduce such chemical signatures. 
Gas-rich mergers as the dominant  process in forming thick disks provide 
a natural explanation of observed abundance patterns and gradients in the 
Milky Way disk components. A high-redshift epoch of gas-rich  mergers is 
emerging as an important phase in the very birth of disk galaxies.

\acknowledgments

The simulation was performed at the Laboratoire d'Astrophysique 
Num\'erique, Universit\'e Laval.
CB, SR, \& HM are funded by the Canada Research Chair program 
and NSERC. DK is a JSPS Fellow.

\fontsize{10}{10pt}\selectfont

\end{document}